\documentclass[prd,aps,floats,preprintnumbers,preprint]{revtex4}
\usepackage{amsmath,amssymb}
\usepackage{eucal}
\usepackage{graphicx}

\begin{document}

\preprint{SU-GP-03/6-1} \preprint{SU-4252-784}
\setlength{\unitlength}{1mm}

\title{The Shapes of Dirichlet Defects}

\author{Mark Bowick\footnote{bowick@physics.syr.edu}, Antonio De
Felice\footnote{defelice@physics.syr.edu} and Mark
Trodden\footnote{trodden@physics.syr.edu}} \affiliation{ Department of
Physics, Syracuse University, Syracuse, NY 13244-1130, USA.  }

\begin{abstract}
If the vacuum manifold of a field theory has the appropriate
topological structure, the theory admits topological structures
analogous to the D-branes of string theory, in which defects of one
dimension terminate on other defects of higher dimension.  The shapes
of such defects are analyzed numerically, with special attention paid
to the intersection regions. Walls (co-dimension 1 branes) terminating
on other walls, global strings (co-dimension 2 branes) and local
strings (including gauge fields) terminating on walls are all
considered. Connections to supersymmetric field theories, string
theory and condensed matter systems are pointed out.
\end{abstract}


\maketitle

\section{Introduction.}
\label{introduction}

Topological defects arise whenever spontaneous symmetry breaking leads
to a topologically nontrivial manifold of vacua or ground
states. Defect configurations involving spatially nonuniform order
fields are of considerable interest in any field theory which
possesses them as their topological quantum numbers distinguish them
from vacua with uniform order fields. A complete understanding of the
physical significance of defects requires a good knowledge of their
field and energy profile and much effort has been devoted over the
years to this subject (see~\cite{VS:94,Hindmarsh:1994re} for
reviews). Applications range all the way from fundamental physics to
cosmology and condensed matter phase transitions.

For a symmetry breaking phase transition $G \rightarrow H$, the vacuum
manifold is given by the coset space $\mathcal M = G/H$, and
$p$-dimensional defects are classified by the homotopy group
$\pi_{d-p-1} (\mathcal M)$, where $d$ is the spatial
dimensionality. For field theories with a relatively simple field
content and abelian homotopy groups the associated topological defects
are essentially isolated fundamental objects {--} they do not
intersect.

In more complicated field theories new possibilities arise involving
composite defects or intersecting arrays of defects.  Such
configurations add a new layer of mathematical and physical
richness. The simplest examples are $p$-dimensional defects whose
boundaries are themselves ($p-1$)-dimensional
defects~\cite{Kibble:1982dd,Vilenkin:hm,Hindmarsh:xc,Chamblin:1999ze}. Another
example is furnished by theories with non-abelian homotopy groups,
such as the quaternionic fundamental group of $RP^2$, in which
topological defects of a fixed dimension (strings) intersect.  Some
time ago Carroll and Trodden~\cite{Carroll:1997pz,Trodden:1998ne}
considered the case of defects which terminate when they intersect
defects of equal or higher dimensionality, such as strings ($p=1$)
ending on domain walls ($p=2$). The defects on which other defects end
were termed {\em Dirichlet topological defects} (DTD), in analogy with
D-branes in string
theory~\cite{Horava:1989vt,Horava:1989ga,Dai:ua,Polchinski:1995mt,Polchinski:1996fm,Polchinski:1996na},
extended objects on which fundamental strings can terminate. It is
important, of course, to keep in mind that there are important
differences between the two sets of objects, as emphasized
in~\cite{Carroll:1997pz,Dvali:2002fi}.

Related configurations are of interest in several contexts.  Strings
terminating on walls can arise in Yang-Mills
theories~\cite{Witten:1997ep,Campos:1998db,Holland:2000uj,Dvali:1999pk}
as well as in grand unified models, where they may be of use in
tackling the monopole problem~\cite{Dvali:1997sa,Alexander:1999mf}. In
nonabelian gauge theories, non-intercommuting cosmic strings provide a
further example. In addition, in higher spatial dimensions, the
required symmetry breaking schemes can lead to a nonabelian unbroken
gauge group, and the phenomenon of confinement for the resulting
defects. The rich variety of symmetry breakings and associated ground
state manifolds in condensed matter systems provides a particularly
concrete context to study topological defects, including the Dirichlet
class studied here. In fact an $N=1$ vortex filament in the nonchiral
superfluid $^3He-B$ phase can terminate at a domain wall interface
separating this phase from a superfluid $^3He-A$
phase~\cite{Volovik:1999sc}, thus realizing a string terminating on a
wall.

Since interesting nonlinearities arise both in the cores of such
defects and at the points at which distinct classes of defects
intersect, it is important to have explicit solutions at hand to
study, rather than simply a classification of the associated vacuum
structure which guarantees their existence. Some work along these
lines has already been performed, particularly in the case of defects
networks in supersymmetric theories\footnote{For microphysical
properties of defects in supersymmetric theories
see~\cite{Davis:1997bs,Davis:1997ny,Morris:at,Morris:1995wd,Morris:1997ua}.}~\cite{Bazeia:1998zv,Saffin:1999au,Bazeia:1999xi,Bazeia:2002xt}
(for related work in the context of D-branes
see~\cite{Witten:1997sc,Callan:1997kz,Gibbons:1997xz,Howe:1997ue,Hashimoto:1997px}).
This is the focus of the current paper, whose outline is now
provided. In Section~\ref{wew} we treat the case of walls ending on
walls, followed in Section~\ref{sew} by the case of global strings
ending on walls. In Section~\ref{gsew} we add gauge fields and treat
the case of local strings terminating on walls. In each case we give
explicit field and energy profiles. Throughout this paper we work in
three spatial dimensions.

\section{Walls ending on walls}
\label{wew}

\subsection{The Model}
The simplest example of a DTD is a wall ending on another wall.  Walls
are produced by the spontaneous breaking of discrete
symmetries. Consider a theory with three fields, $\phi$, $\psi_1$ and
$\psi_2$, invariant under three $Z_2$ symmetries. Such configurations
are a special case of $Z_n$ walls~\cite{Ryden:1989vj}, and are related
to solutions in
supersymmetric~\cite{Gibbons:1999np,Carroll:1999wr,Gauntlett:2000de}
and supergravity~\cite{Carroll:1999mu,Hashimoto:1998ug} theories.
\begin{alignat}8
&Z^{(1)}_2:{}&&(&&\phi  &&\mapsto-\phi,\; \psi_1\leftrightarrow\psi_2)\notag\\
&Z^{(2)}_2:{}&& &&\psi_1&&\mapsto -\psi_1\\
&Z^{(3)}_2:{}&& &&\psi_2&&\mapsto -\psi_2 \ , \notag
\end{alignat}
with Lagrangian density~\cite{Carroll:1997pz}
\begin{equation}
\mathcal{L} = \tfrac12\,\partial_\mu\phi\,\partial^\mu\phi+
\tfrac12\,\partial_\mu\psi_1\,\partial^\mu\psi_1+
\tfrac12\,\partial_\mu\psi_2\,\partial^\mu\psi_2-V(\phi,\psi_1,\psi_2) \ ,
\end{equation}
where we choose the most general renormalizable potential
invariant under $Z_2^{(1)}\times Z_2^{(2)}\times Z_2^{(3)}$
\begin{equation}
V(\phi,\psi_1,\psi_2) = \lambda_\phi(\phi^2-\tilde v^2)^2+
    \lambda_\psi\,[\psi_1^2+\psi_2^2-\tilde w^2+
    g(\phi^2-\tilde v^2)]^2+h\psi_1^2\psi_2^2-
    \mu\phi(\psi_1^2-\psi_2^2) \ .
\end{equation}
For $\mu=0$, the minima of $V$ are the 8 points given by
\begin{equation}
\phi=\pm\tilde v\quad\text{and}\quad
\begin{cases}
(\psi_1,\psi_2)=(\pm\tilde w,0)\\
\text{or}\\
(\psi_1,\psi_2)=(0,\pm\tilde w) \ .
\end{cases}
\end{equation}
Since the potential is positive-definite in this case, the presence of
$h$ requires either $\psi_1$ or $\psi_2$ to vanish, after which the
first term in the potential is minimized by $\phi=\pm\tilde v$ and the
second term by $\psi_1=\pm\tilde w$ or $\psi_2=\pm\tilde w$.  For
$\mu>0$ the degeneracy of the 8 vacua is broken, leaving only 4 minima
of the potential, at
\begin{alignat}6
\phi &=   &&v, \qquad&(\psi_1,\psi_2) &= (\pm w,   0)\ ;\\
\phi &=-  &&v,       &(\psi_1,\psi_2) &= (0,\pm w) \ ,
\end{alignat}
where $v$ is the solution of the cubic equation
\begin{equation}
8\lambda_\phi\lambda_\psi\,v^3+6\lambda_\psi g \mu\,v^2-
(8\lambda_\phi\lambda_\psi\tilde v^2+\mu^2)\,v-
2\lambda_\psi\mu\,(g\tilde v^2+\tilde w^2)=0 \ ,
\end{equation}
satisfying $\displaystyle\lim_{\mu\to 0} v = \tilde v$, and $w$ is
given by
\begin{equation}
w = \left[\tilde w^2+g(\tilde v^2-v^2)+\frac{\mu v}{2\lambda_\psi}
\right]^{\!1/2} \ .
\end{equation}

The equations of motion for static fields are then
\begin{alignat}{6}
&\nabla^2\phi  &&=
    4\lambda_\phi\phi(\phi^2-\tilde v^2)+
    4g\lambda_\psi\phi[\psi_1^2+\psi_2^2-\tilde
    w^2+g(\phi^2-\tilde v^2)]-\mu(\psi_1^2-\psi_2^2) \label{eq_wphi}\\
&\nabla^2\psi_1 &&=
    4\lambda_\psi\psi_1[\psi_1^2+\psi_2^2-\tilde
    w^2+g(\phi^2-\tilde v^2)]+2h\psi_1\psi_2^2
    -2\mu\phi\psi_1 \label{eq_wpsi1}\\
&\nabla^2\psi_2 &&=
    4\lambda_\psi\psi_2[\psi_1^2+\psi_2^2-\tilde
    w^2+g(\phi^2-\tilde v^2)]+2h\psi_1^2\psi_2
    +2\mu\phi\psi_2 \ . \label{eq_wpsi2}
\end{alignat}

This model is constructed to admit solutions consisting of domain
walls terminating on other domain walls. Such a configuration is
intrinsically two-dimensional, and hence is independent of one
spatial coordinate, which we choose to be $z$. What remains is a
boundary value problem for an elliptic system of partial
differential equations, for which we must prescribe the value of
the fields on the boundary.

\subsection{The Numerical Problem - Boundary Conditions}
\begin{figure}[ht]
\centering
\includegraphics[width=2in]{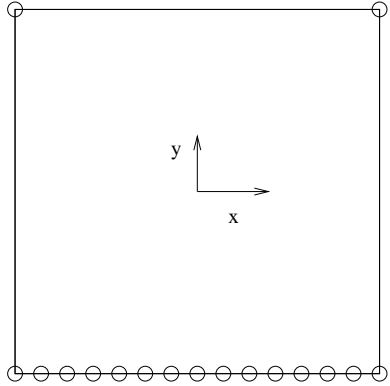}
\caption{Boundary conditions for walls on walls. The circles
($\circ$) represent Dirichlet boundary conditions and solid
lines represents Neumann boundary conditions.} \label{bd_ww}
\end{figure}

Our choice of boundary conditions is illustrated in
Fig.~\ref{bd_ww},
so that we have Dirichlet boundary conditions
\begin{align}
\phi(  &x,y=-\infty)=-v\ ,\notag\\
\psi_1(&x,y=-\infty)=0\ , \\
\psi_2(&x,y=-\infty)=w \ , \notag
\end{align}
at $y\rightarrow -\infty$, and Neumann boundary
conditions everywhere else except at the upper corners, at which we choose
\begin{align}
\phi(  &x=\pm\infty,y=+\infty)=v \ ,\notag\\
\psi_1(&x=\pm\infty,y=+\infty)=\pm w \ ,\\
\psi_2(&x=\pm\infty,y=+\infty)=0 \ . \notag
\end{align}
These choices define a wall in the $\psi_1$ field, terminating on
a wall in the $\phi$ field {--} the Dirichlet wall.

The numerical algorithms we use require homogeneous boundary
conditions and so we introduce new fields
\begin{alignat}4
&\phi  &&=u_0+F_0\ ,\notag\\
&\psi_1&&=u_1+F_1 \ ,\\
&\psi_2&&=u_2+F_2 \ ,\notag
\end{alignat}
and define
\begin{alignat}4
&F_0(x,y)&&=f_{0s}(y)+\frac{1+\tanh(2y)}2\,
  \left[f_{0u}(x)-f_{0s}(y=+\infty)\right] \ ,\notag\\
&F_1(x,y)&&=f_{1s}(y)\tanh(2x)+\frac{1+\tanh(2y)}2\,
  \left[f_{1u}(x)-f_{1s}(y=+\infty)\tanh(2x)\right] \ ,\\
&F_2(x,y)&&=f_{2s}(y) \ . \notag
\end{alignat}
The functions $f_{iu}(x)$ and $f_{is}(y)$ are the solutions of the 
systems of ordinary differential equations obtained by restricting 
(\ref{eq_wphi}-\ref{eq_wpsi2}) to the ``up'' boundary
($y=+\infty$) and the ``side'' boundary ($x=+\infty$) respectively, 
with $i=0,1,2$ corresponding to $\phi,\psi_1,\psi_2$ respectively.

These functions are easily found using standard algorithms 
with boundary conditions
\begin{align}
f_{0u}(\pm\infty)=& v\ ,\notag\\
f_{1u}(\pm\infty)=&\pm w \ ,\\
f_{2u}(\pm\infty)=&0 \ . \notag
\end{align}
\begin{alignat}4
f_{0s}(-\infty)&=-v\qquad&f_{0s}(+\infty) &= v\notag\\
f_{1s}(-\infty)&=0 \qquad&f_{1s}(+\infty) &= w\\
f_{2s}(-\infty)&=w \qquad&f_{2s}(+\infty) &= 0 \ .\notag
\end{alignat}
The symmetries of our problem are such that the solutions on the
boundary at $x=-\infty$ are then $\psi_1(x=-\infty,y) = -f_{1s}(y)$
and $\psi_2(x=-\infty,y)=f_{2s}(y)$.  Note that one solution
$f_{2u}(x)\equiv 0$ is trivial.

We are now in a position to solve numerically our partial differential
equations in the interior. We use a multigrid
algorithm~\cite{Multigrid} to solve for the fields $u_i$, satisfying
\begin{alignat}3
u_i &=0&\qquad&\text{at the $y=-\infty$ boundary and at $(x,y)
=(\pm\infty,+\infty)$ \ ,}\notag\\ \partial_x u_i &=0&\qquad&\text{at
$x=\pm\infty$ \ ,}\\ \partial_y u_i &=0&\qquad&\text{at $y=+\infty$}\
.\notag
\end{alignat}

\subsection{Results - Walls on Dirichlet Walls}
We present our results as a set of plots of field profiles and energy
densities, providing a number of ways of viewing  configuration shapes.
We have already defined the field profiles, and the
energy density for static fields is given by
\begin{equation}
\begin{split}
E = T^0{}_0 = -\mathcal{L} &=
\frac12\left(\frac{\partial\phi}{\partial x}\right)^{\!2}+
\frac12\left(\frac{\partial\phi}{\partial y}\right)^{\!2}+
\frac12\left(\frac{\partial\psi_1}{\partial x}\right)^{\!2}+
\frac12\left(\frac{\partial\psi_1}{\partial y}\right)^{\!2}+\\
&\qquad \frac12\left(\frac{\partial\psi_2}{\partial
x}\right)^{\!2}+ \frac12\left(\frac{\partial\psi_2}{\partial
y}\right)^{\!2}+ V(\phi,\psi_1,\psi_2) \ .
\end{split}
\end{equation}

\begin{figure}[ht]
\centering \includegraphics[width=4.5in]{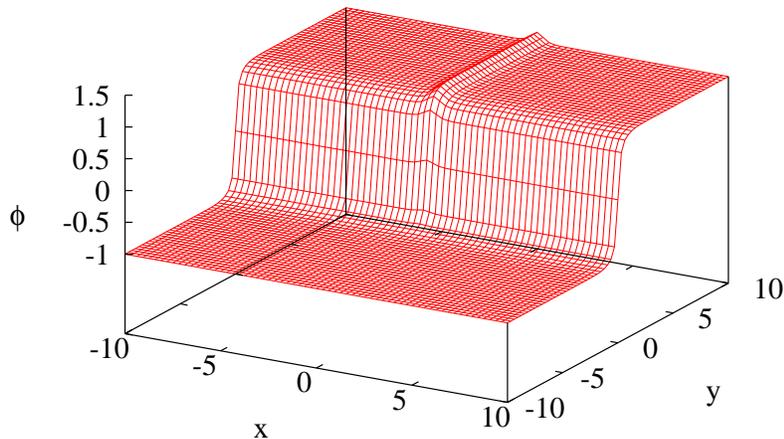}
\caption{The $\phi$ Dirichlet wall configuration.} \label{ww_phi}
\end{figure}

\begin{figure}[ht]
\centering
\includegraphics[width=4.5in]{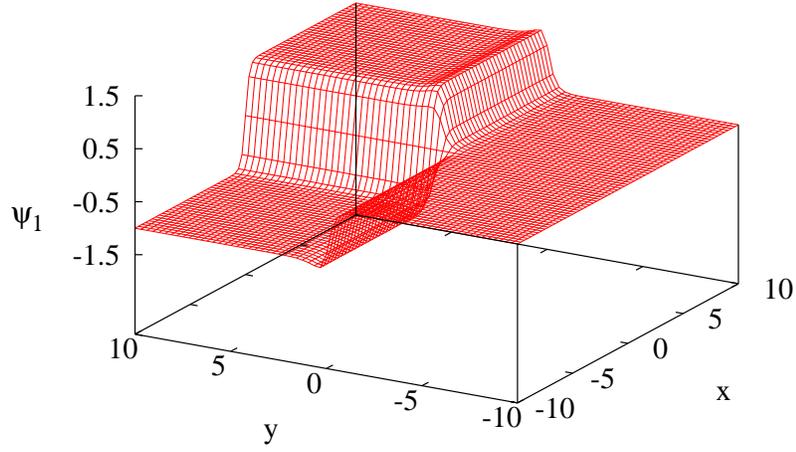}
\caption{Field configuration for $\psi_1$.} \label{ww_psi1}
\end{figure}

\begin{figure}[ht]
\centering
\includegraphics[width=4.5in]{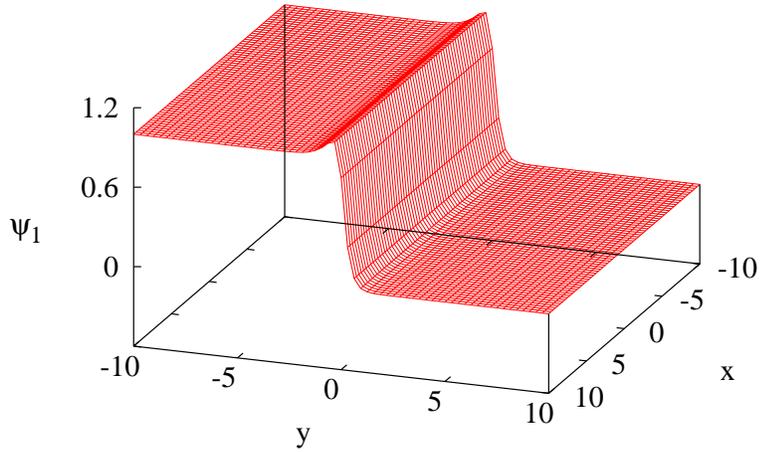}
\caption{Walls on walls field configuration for $\psi_2$.}
\label{ww_psi2}
\end{figure}

\begin{figure}[ht]
\centering
\includegraphics[width=4.5in]{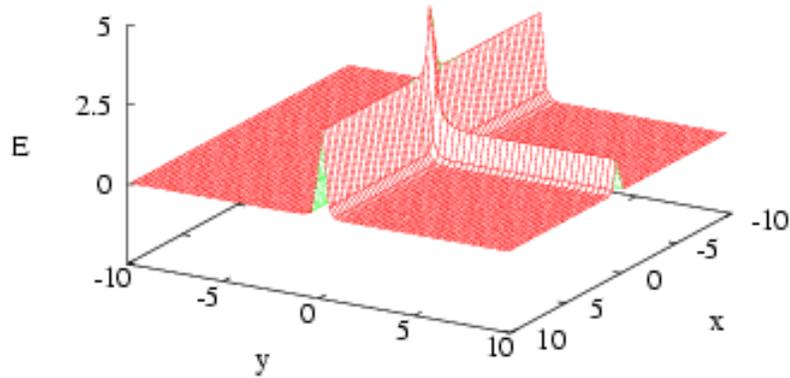}
\caption{Energy density for walls ending on walls.}\label{ww_en}
\end{figure}

\begin{figure}[ht]
\centering
\includegraphics[width=4.5in]{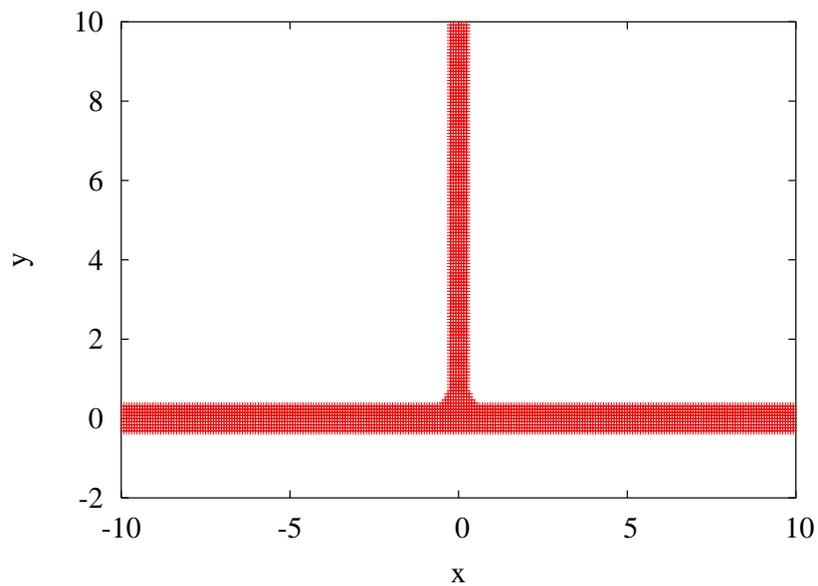}
\caption{Contour plot of the energy density for walls ending on
walls.} \label{ww}
\end{figure}

The $\phi$ field domain wall configuration is plotted in
Fig.~\ref{ww_phi}. The intersection with the $\psi_1$ wall at
$x=0$ is clearly visible.
The corresponding configuration for $\psi_1$ is shown in
Fig.~\ref{ww_psi1}. Note the domain wall for $y=+\infty$.
Finally the $\psi_2$ field configuration is plotted in
Fig.~\ref{ww_psi2}. It has the form of a wall with half the height of
the Dirichlet wall in the sense that it goes from 0 (at $y=+\infty$)
to $w=1$ (at $y=-\infty$).
The energy density plotted in Fig.~\ref{ww_en} illustrates the merging
of the walls at the origin.
A corresponding contour plot for the energy density in
Fig.~\ref{ww}  shows the wall merging very clearly.

\section{Global strings ending on walls}
\label{sew}
\subsection{The Model}
In the simplest case, strings are generated by a breakdown of a $U(1)$
symmetry. In analogy with our approach in the previous section,
consider the following transformations on one real scalar field $\phi$
and two complex scalar fields $\psi_1$ and $\psi_2$.
\begin{alignat}4
Z_2:{} &(\phi &&\mapsto-\phi, \psi_1\leftrightarrow\psi_2)\notag\\
U(1)_1:{}&\psi_1&&\mapsto {\rm e}^{-i\omega_1}\,\psi_1\\
U(1)_2:{}&\psi_2&&\mapsto {\rm e}^{-i\omega_2}\,\psi_2 \ . \notag
\end{alignat}
In this section, both $U(1)$ symmetries are global, i.e.\ $\omega_i$
is space-time independent. The Lagrangian density is
\begin{equation}
\mathcal{L} = \tfrac12\,\partial_\mu\phi\,\partial^\mu\phi+
\partial_\mu\bar\psi_1\,\partial^\mu\psi_1+
\partial_\mu\bar\psi_2\,\partial^\mu\psi_2-V(\phi,\psi_1,\psi_2) \ ,
\end{equation}
with potential
\begin{eqnarray}
V(\phi,\psi_1,\psi_2) = \lambda_\phi(\phi^2-\tilde v^2)^2 &+&
    \lambda_\psi\,\bigl[|\psi_1|^2+|\psi_2|^2-\tilde w^2+
    g(\phi^2-\tilde v^2)\bigr]^2 \nonumber \\
&+& h|\psi_1|^2|\psi_2|^2-
    \mu\phi(|\psi_1|^2-|\psi_2|^2) \ .
\end{eqnarray}

Consider the following static ansatz
\begin{alignat}4
&\phi  &&=\phi(r,z)\notag\\
&\psi_1&&=R_1(r,z)\,e^{i\theta}\\
&\psi_2&&=\psi_2(r,z) \ ,\notag
\end{alignat}
where $\phi$, $R_1$ and $\psi_2$ are real functions, for which the
equations of motion are
\begin{alignat}4
&\nabla^2\phi  &&= 4\lambda_\phi\phi(\phi^2-\tilde v^2)+
  4\lambda_\psi g \phi[R_1^2+\psi_2^2-\tilde w^2+g(\phi^2-\tilde v^2)]
  -\mu(R_1^2-\psi_2^2) \ ,  \label{eq_gphi} \\
&\nabla^2R_1 &&= 2\lambda_\psi R_1\,[R_1^2+\psi_2^2-\tilde w^2+g(\phi^2-\tilde v^2)]
  +hR_1\psi_2^2-\mu\phi R_1 \ ,  \label{eq_gR1} \\
&\nabla^2\psi_2 &&= 2\lambda_\psi\psi_2\,[R_1^2+\psi_2^2-\tilde w^2+g(\phi^2-\tilde v^2)]
  +hR_1^2\psi_2+\mu\phi\psi_2 \ .
\label{eq_gpsi2}
\end{alignat}
\subsection{The Numerical Problem - Boundary Conditions}
\begin{figure}[ht]
\centering
\includegraphics[width=1.75in]{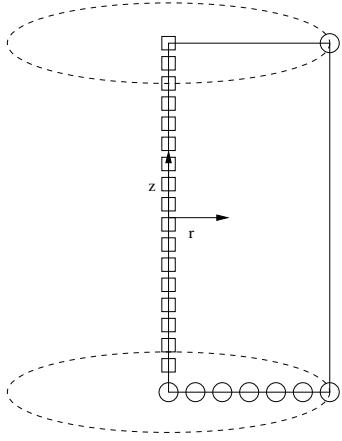}
\caption{Boundary conditions for a string ending on a wall. The
circles ($\circ$) represent Dirichlet boundary conditions, the
squares ($\Box$) represent mixed boundary conditions (Dirichlet
for some fields, Neumann for the remaining ones) and solid lines
represent Neumann boundary conditions.} \label{bd_sw}
\end{figure}

We now proceed as in the previous section, imposing the boundary
conditions illustrated in Fig.~\ref{bd_sw}.

We choose Dirichlet boundary conditions for the bottom
boundary
\begin{align}
\phi(  &r,z=-\infty)=-v \ ,\notag\\
R_1(   &r,z=-\infty)=0 \ , \\
\psi_2(&r,z=-\infty)=w \ ,\notag
\end{align}
and at the upper corner we impose
\begin{align}
\phi(  &r=\infty,\,z=+\infty)=v \ ,\notag\\
R_1(   &r=\infty,\,z=+\infty)=w \ ,\\
\psi_2(&r=\infty,\,z=+\infty)=0 \ .\notag
\end{align}

The boundary $r=0$ requires a more detailed discussion. Taylor
expanding the fields around $r=0$, we have
\begin{equation}
\phi  =\sum_n a_n(z)\,r^n \ ,\qquad
R_1   =\sum_n b_n(z)\,r^n \ ,\qquad
\psi_2=\sum_n c_n(z)\,r^n \ .
\end{equation}
Equations [\ref{eq_gphi}-\ref{eq_gpsi2}], in the limit $r\to0$, become
\begin{alignat}8
&\frac{a_1(z)}r&&+4a_2(z)
&&+\frac{d^2a_0(z)}{dz^2}&&=f_0[a_0(z),b_0(z),c_0(z)] \ ,\notag\\
&3b_2(z)
&&-\frac{b_0(z)}{r^2}&&+\frac{d^2b_0(z)}{dz^2}&&=f_1[a_0(z),b_0(z),c_0(z)]
\ ,\\
&\frac{c_1(z)}r&&+4c_2(z)
&&+\frac{d^2c_0(z)}{dz^2}&&=f_2[a_0(z),b_0(z),c_0(z)] \ . \notag
\end{alignat}
Demanding regularity at the origin gives
\begin{alignat}4
a_1(z)&=0\qquad&\left.\partial_r\phi\right|_{r=0}&=0 \ ,\notag\\
b_0(z)&=0\qquad&\left.R_1\right|_{r=0}&=0 \ ,\\
c_1(z)&=0\qquad&\left.\partial_r\psi_2\right|_{r=0}&=0 \ .\notag
\end{alignat}
For the other boundaries we imposed Neumann boundary conditions for
all the fields.

As for the case of walls on walls, we define new fields with
homogeneous boundary conditions. Defining the ``up'' boundary to be at
$z=+\infty$ and the ``side'' boundary to be at $r=+\infty$ we again
introduce the functions $f_{iu}(r)$ and $f_{is}(z)$ which are
solutions to the appropriate systems of ordinary differential
equations with boundary conditions
\begin{alignat}4
\partial_r f_{0u}(0)=0&\qquad&f_{0u}(+\infty) &= v \ ,\notag\\
f_{1u}(0)=0&\qquad&f_{1u}(+\infty)            &= w \ ,\\
\partial_r f_{2u}(0)=0&\qquad&f_{2u}(+\infty) &= 0 \ ,\notag
\end{alignat}
\begin{alignat}4
f_{0s}(-\infty)&=-v\qquad&f_{0s}(+\infty) &= v \ ,\notag\\
f_{1s}(-\infty)&=0 \qquad&f_{1s}(+\infty) &= w \ ,\\
f_{2s}(-\infty)&=w \qquad&f_{2s}(+\infty) &= 0 \ .\notag
\end{alignat}
Again we have a trivial solution $f_{2u}(r) \equiv 0$.  
Following the previous section, we define the following
functions
\begin{alignat}4
&F_0(r,z)&&=f_{0s}(z)+\frac{1+\tanh(2z)}2\,
  \left[f_{0u}(r)-f_{0s}(z=+\infty)\right] \ ,\notag\\
&F_1(r,z)&&=f_{1s}(z)\tanh(r)+\frac{1+\tanh(2z)}2\,
  \left[f_{1u}(r)-f_{1s}(z=+\infty)\tanh(r)\right] \ ,\\
&F_2(r,z)&&=f_{2s}(z) \ , \notag
\end{alignat}
and introduce new fields
\begin{alignat}6
&\phi  &&=u_0&&+F_0 \ ,\notag\\
&\psi_1&&=u_1&&+F_1 \ ,\\
&\psi_2&&=u_2&&+F_2 \ ,\notag
\end{alignat}
obeying homogeneous boundary conditions
\begin{alignat}4
u_i &=0&\qquad&\text{at the $z=-\infty$ boundary, at
$(r,z)=(+\infty,+\infty)$} \notag \\ &&\qquad&\text{and for $u_1$ at
the core of the string,}\notag\\ \partial_r u_i &=0&\qquad&\text{at
the $r=+\infty$ boundary and for $i=0,2$ at the core of the
string,}\notag\\ \partial_z u_i &=0&\qquad&\text{at the $z=+\infty$
boundary} \ .\notag
\end{alignat}

\subsection{Results - Global Strings Ending on Dirichlet Walls}
\begin{figure}[ht]
\centering
\includegraphics[width=4.5in]{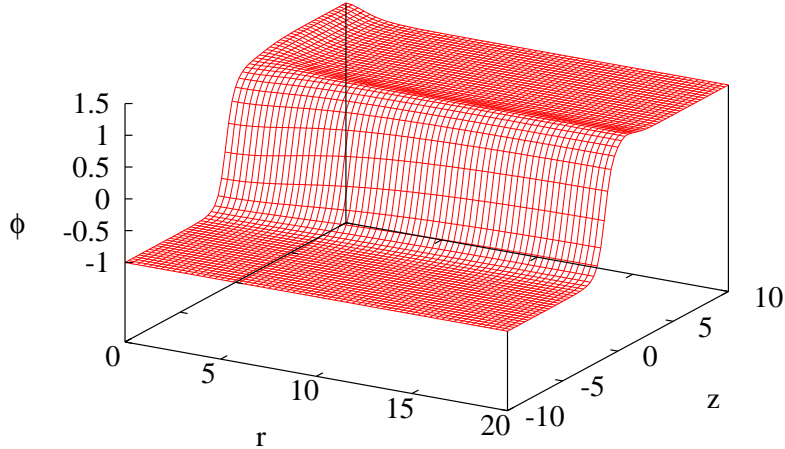}
\caption{Field $\phi$, the Dirichlet wall.} \label{sw_phi}
\end{figure}

\begin{figure}[ht]
\centering
\includegraphics[width=4.2in]{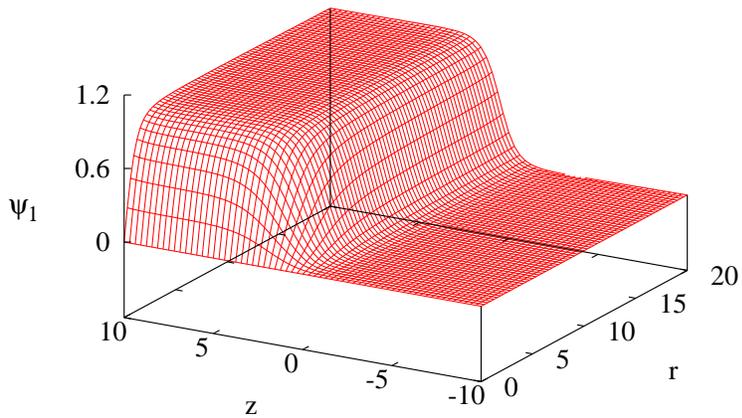}
\caption{Field $\psi_1$, the global string.} \label{sw_psi1}
\end{figure}

\begin{figure}[ht]
\centering
\includegraphics[width=4.2in]{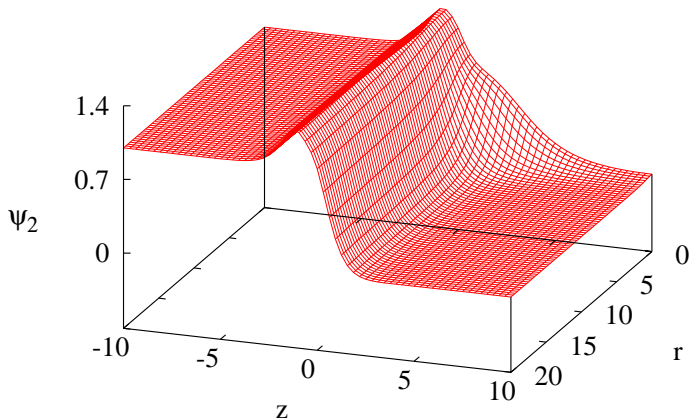}
\caption{Field $\psi_2$ for a global string and a wall.}
\label{sw_psi2}
\end{figure}

To help in understanding our plots, note that the energy density for
static fields in this section is given by
\begin{equation}
\begin{split}
E = T^0{}_0 = -\mathcal{L} &=
\frac12\left(\frac{\partial\phi}{\partial r}\right)^{\!2}+
\frac12\left(\frac{\partial\phi}{\partial z}\right)^{\!2}+
\left(\frac{\partial\psi_2}{\partial r}\right)^{\!2}+
\left(\frac{\partial\psi_2}{\partial z}\right)^{\!2}+\\
&\qquad
\left(\frac{\partial R_1}{\partial r}\right)^{\!2}+
\frac{R_1^2}{r^2} + \left(\frac{\partial R_1}{\partial z}\right)^{\!2}+
V(\phi,R_1,\psi_2) \ .
\end{split}
\end{equation}
The $\phi$ field domain wall configuration is plotted in
Fig.~\ref{sw_phi}. The intersection with the $\psi_1$ string at $r=0$
is clearly visible.
The corresponding configuration for $\psi_1$ is shown in
Fig.~\ref{sw_psi1}. Note the string for $z=+\infty$.
Finally the $\psi_2$ field configuration is plotted in
Fig.~\ref{sw_psi2}. It has the form of a wall (except for an
interaction with the string for $r=0$ and $z>0$) with half the
height of the Dirichlet wall in the sense that it goes from 0 (at
$z=+\infty$) to $w=1$ (at $z=-\infty$).

\begin{figure}[ht]
\centering
\includegraphics[width=4.5in]{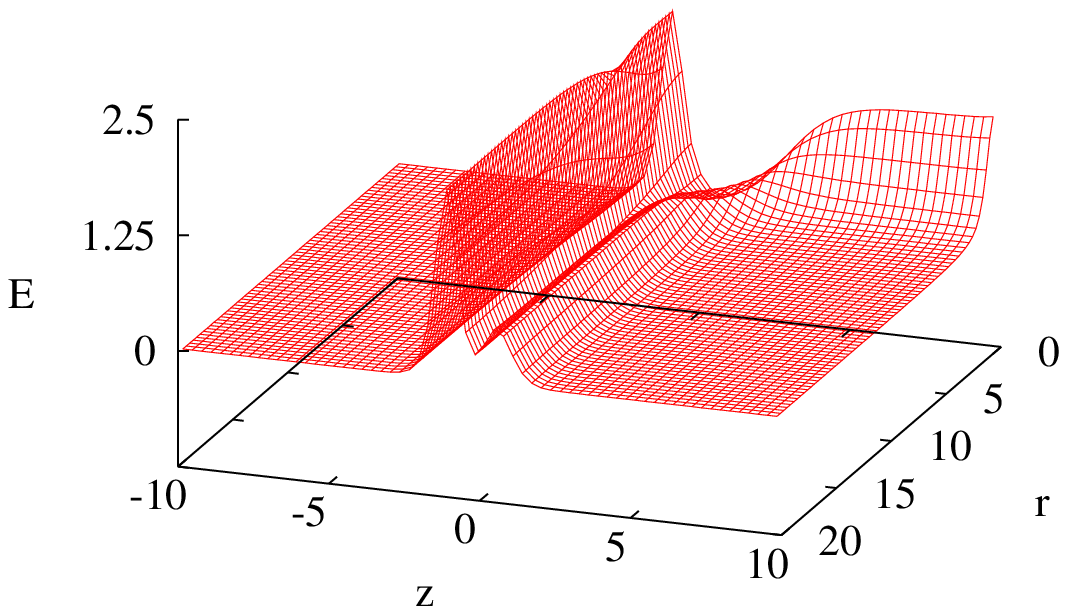}
\caption{Energy density for a global string ending on a wall.}
\label{sw_en}
\end{figure}

\begin{figure}[ht]
\centering
\includegraphics[width=4.5in]{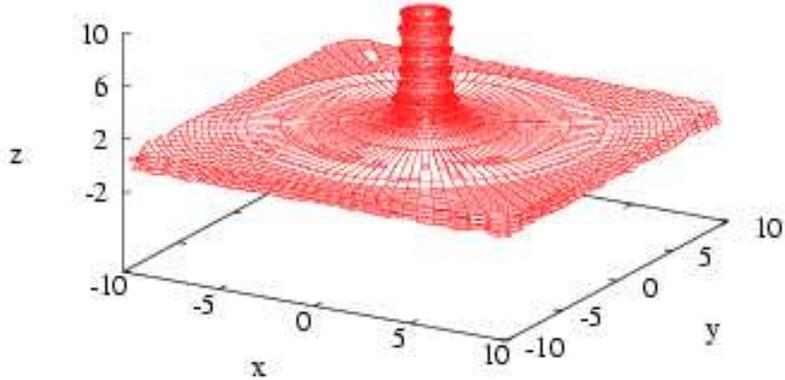}
\caption{Contour plot of the energy density for a global string
ending on a wall.} \label{sw}
\end{figure}

The energy density plotted in Fig.~\ref{sw_en} illustrates the
merging of the string with the wall at the origin.
A corresponding contour plot for the energy density in
Fig.~\ref{sw} shows the string ending on the wall very clearly.

\section{Gauge strings on walls}
\label{gsew} More interesting connections with physical models can
be made by introducing gauge fields. In this section we take the
first steps towards this by finding numerical solutions for local
cosmic strings terminating on Dirichlet domain walls.

\section{The Model}
We modify the Lagrangian density of the last section by introducing
a $U(1)$ gauge field $A_{\mu}$, under which the field $\psi_1$ is charged.
\begin{equation}
\mathcal{L} = \tfrac12\,\partial_\mu\phi\,\partial^\mu\phi+
\bar D_\mu\bar\psi_1\,D^\mu\psi_1+
\partial_\mu\bar\psi_2\,\partial^\mu\psi_2-
\tfrac14\,F^{\mu\nu}F_{\mu\nu}-V(\phi,\psi_1,\psi_2) \ ,
\end{equation}
where
\begin{equation}
D_\mu = \partial_\mu - ie A_\mu \ .
\end{equation}
The equations of motion for $\phi$ and $\psi_2$ are unchanged from
the case of the global string, since we have constructed the model
so that the relevant winding occurs for the $\psi_1$ field only.
Thus, the modified equation of motion for $\psi_1$ is
\begin{equation}
\label{primapsi1}
\partial_\mu\partial^\mu\psi_1-ie\partial_\mu(A^\mu\psi_1)+
\frac{\partial V}{\partial\bar\psi_1}-
ieA_\mu(\partial^\mu\psi_1-ieA^\mu\psi_1)=0 \ .
\end{equation}
Imposing the Lorentz condition $\partial_{\mu} A^{\mu}=0$ for a static gauge field yields
\begin{equation}
\vec\nabla\cdot\vec A = 0 \ ,
\end{equation}
and equation (\ref{primapsi1}) then becomes
\begin{equation}
\label{secondapsi1}
\nabla^2\psi_1=\frac{\partial V}{\partial\bar\psi_1}-
2ie\vec A\cdot\vec\nabla\psi_1+e^2(\vec A\cdot\vec A)\psi_1 \ .
\end{equation}
We are interested in the ansatz
\begin{alignat}4
&\phi  &&=\phi(r,z) \ ,\notag\\
&\psi_1&&=R_1(r,z)\,e^{i\theta} \ ,\\
&\psi_2&&=\psi_2(r,z) \ ,\notag
\end{alignat}
where $\phi$, $R_1$ and $\psi_2$ are real functions.
As is well known for cosmic strings,
the imaginary part of equation (\ref{secondapsi1}) gives
\begin{equation}
A_r\,\frac{\partial R_1}{\partial r}+A_z\,\frac{\partial R_1}{\partial z}=0 \ ,
\end{equation}
with solution
\begin{equation}
\vec A=A\,\hat\theta \ ,
\end{equation}
so that the Lorentz condition becomes $\partial_\theta A=0.$
The equation for $R_1$ then becomes
\begin{alignat}4
&\nabla^2R_1   &&= \frac{\partial^2R_1}{\partial r^2} +
  \frac1r\,\frac{\partial R}{\partial r}-\frac{R_1}{r^2}+
  \frac{\partial^2R_1}{\partial z^2} = \notag\\
&&&=2\lambda_\psi R_1\,[R_1^2+\psi_2^2-\tilde w^2+g(\phi^2-\tilde v^2)]
  +hR_1\psi_2^2-\mu\phi R_1+2e\,\frac{AR_1}r+e^2A^2R_1 \ ,
\end{alignat}
and the equation for the gauge field is
\begin{equation}
\Box A^\alpha = j^\alpha = 2e\,{\rm Im}[\bar\psi_1(\partial^\alpha
-ieA^\alpha)\psi_1] \ ,
\end{equation}
which, for our ansatz, is
\begin{equation}
(\vec\nabla^2\vec A)_\theta =
2e\,R_1^2\,(r^{-1}+eA) \ .
\end{equation}
\subsection{The Numerical Problem - Boundary Conditions}
The geometry of the problem is identical to that for global
strings, and so we refer the reader again to Fig.~\ref{bd_sw}. We
choose Dirichlet boundary conditions for the bottom boundary
\begin{equation}
\begin{split}
\phi(  &r,z=-\infty)=-v \ ,\\
R_1(   &r,z=-\infty)=0 \ , \\
\psi_2(&r,y=-\infty)=w \ ,\\
A(     &r,y=-\infty)=0 \ ,
\end{split}
\end{equation}
and at the upper corner
\begin{equation}
\begin{split}
\phi(  &r=\infty,z=+\infty)=v \ ,\\
R_1(   &r=\infty,z=+\infty)=w \ ,\\
\psi_2(&r=\infty,z=+\infty)=0 \ ,\\
A(     &r=\infty,z=+\infty)=0 \ .
\end{split}
\end{equation}
At the core of the string we choose
\begin{equation}
\begin{split}
\partial_r\phi(r=0,z)&=\partial_r\psi_2(r=0,z) = 0 \ ,\\
R_1(r=0,z) &= A(r=0,z)=0 \ ,
\end{split}
\end{equation}
and for the boundary at $r=+\infty$ we impose Neumann conditions for all
the fields. The choice of these boundary conditions defines a
string ending on a Dirichlet wall.

We again solve ordinary differential equations on the boundaries at
$z=+\infty$ and $r=+\infty$. We keep the same notation as in the
previous section, but now include the gauge field $A$ by defining
$f_{3u}(r)$ and $f_{3s}(z)$ with boundary conditions
\begin{equation}
f_{3u}(0)=0\qquad\partial_r f_{3u}(+\infty)=0 \ ,
\end{equation}
\begin{equation}
f_{3s}(-\infty)=0 \qquad f_{3s}(+\infty)=0 \ .
\end{equation}
Once again $f_{2u}(r)\equiv 0$ and $f_{3s}(z)\equiv 0$ are trivial
solutions.

As before we define
\begin{equation}
\begin{split}
F_0(r,z)&=f_{0s}(z)+\frac{1+\tanh(2z)}2\,
  \left[f_{0u}(r)-f_{0s}(z=+\infty)\right] \ ,\\
F_1(r,z)&=f_{1s}(z)\tanh(r)+\frac{1+\tanh(2z)}2\,
  \left[f_{1u}(r)-f_{1s}(z=+\infty)\tanh(r)\right] \ ,\\
F_2(r,z)&=f_{2s}(z) \ ,\\
F_3(r,z)&=\frac{1+\tanh(2z)}2\,f_{3u}(r) \ ,
\end{split}
\end{equation}
and new fields
\begin{equation}
\begin{split}
\phi  &=u_0+F_0 \ ,\\
\psi_1&=u_1+F_1 \ ,\\
\psi_2&=u_2+F_2 \ ,\\
A     &=u_3+F_3 \ ,
\end{split}
\end{equation}
with homogeneous boundary conditions
\begin{alignat}4
u_i &=0&\qquad&\text{at the $z=-\infty$ boundary, at the corner
  $(r,z)=(+\infty,+\infty)$}\notag \\ &&\qquad&\text{and, for $i=1,3$,
  at the core of the string,}\notag\\ \partial_r u_i
  &=0&\qquad&\text{at the $r=+\infty$ boundary and, for $i=0,2$, at
  the core of the string,}\notag\\ \partial_z u_i &=0&\qquad&\text{at
  the $z=+\infty$ boundary} \ .\notag
\end{alignat}
\subsection{Results - Local Strings Ending on Dirichlet Walls}
\begin{figure}[ht]
\centering
\includegraphics[width=4.5in]{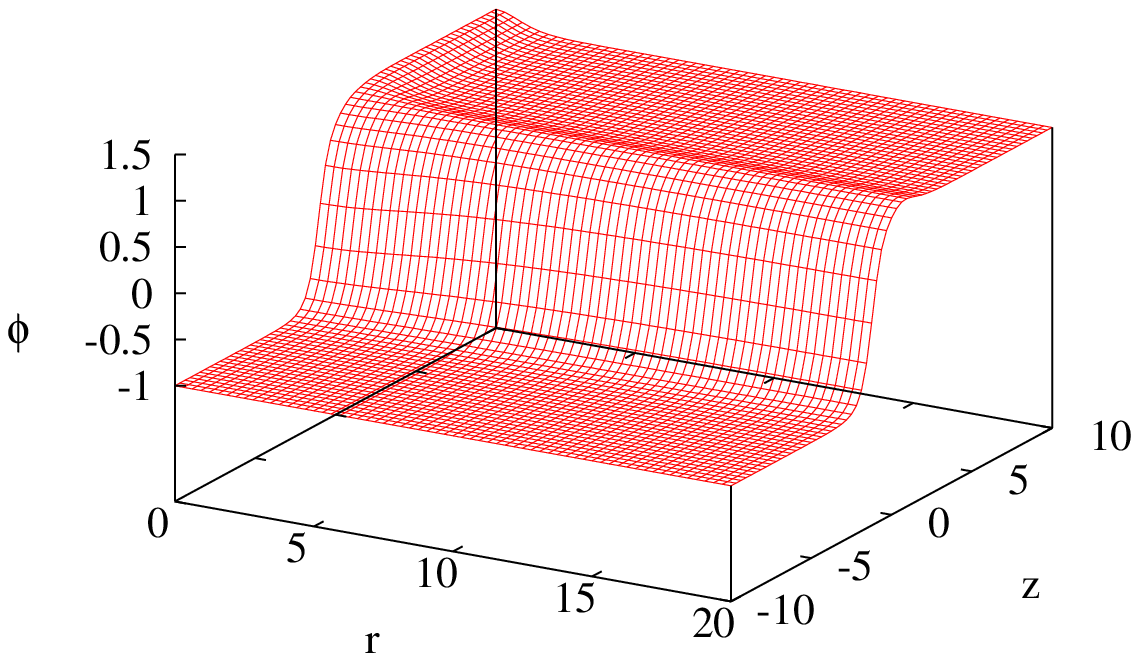}
\caption{The $\phi$ field configuration: Dirichlet wall for local
strings.} \label{lsw_phi}
\end{figure}

\begin{figure}[ht]
\centering
\includegraphics[width=4.25in]{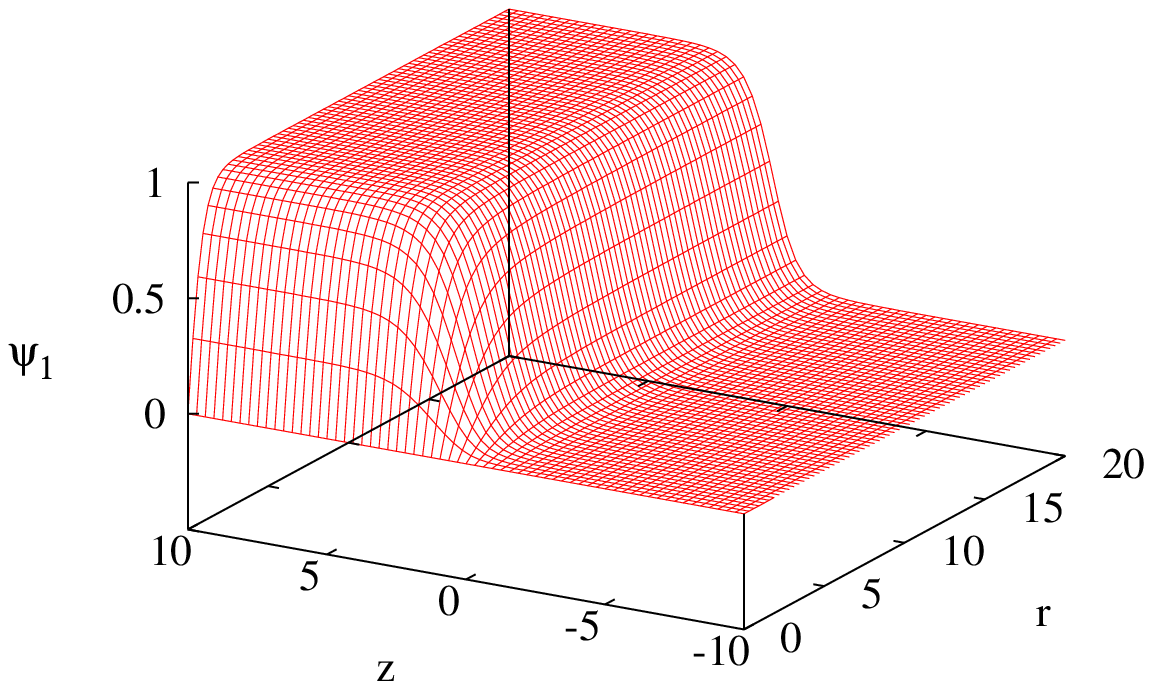}
\caption{The local string field $\psi_1$.} \label{lsw_psi1}
\end{figure}

\begin{figure}[ht]
\centering
\includegraphics[width=4.25in]{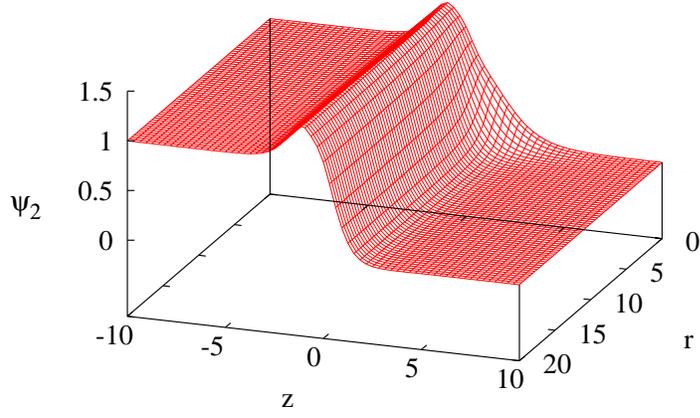}
\caption{Field configuration $\psi_2$ for a local string ending on
a wall.} \label{lsw_psi2}
\end{figure}

\begin{figure}[ht]
\centering
\includegraphics[width=4.5in]{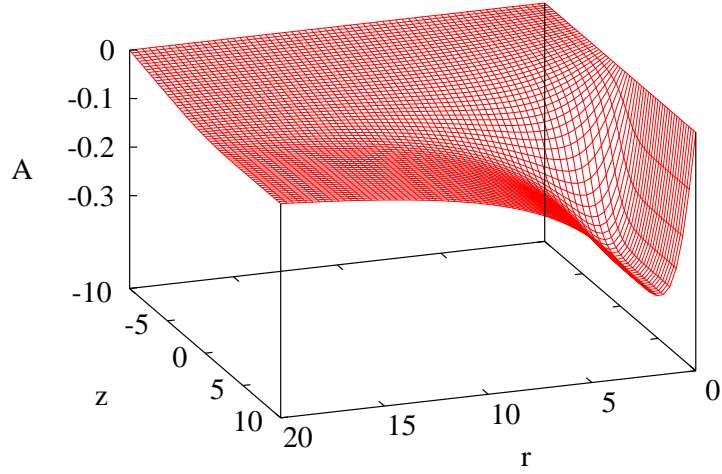}
\caption{The gauge field A.} \label{A}
\end{figure}

\begin{figure}[ht]
\centering
\includegraphics[width=4.5in]{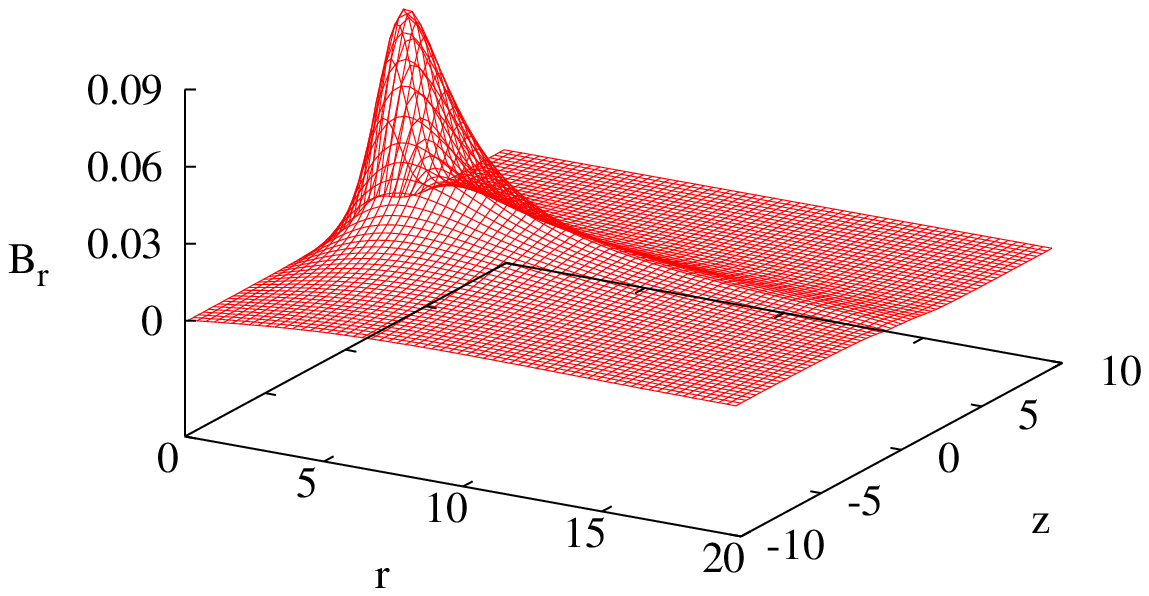}
\caption{Component $B_r$ of the magnetic field.}
\label{Br}
\end{figure}

\begin{figure}[ht]
\centering
\includegraphics[width=4.5in]{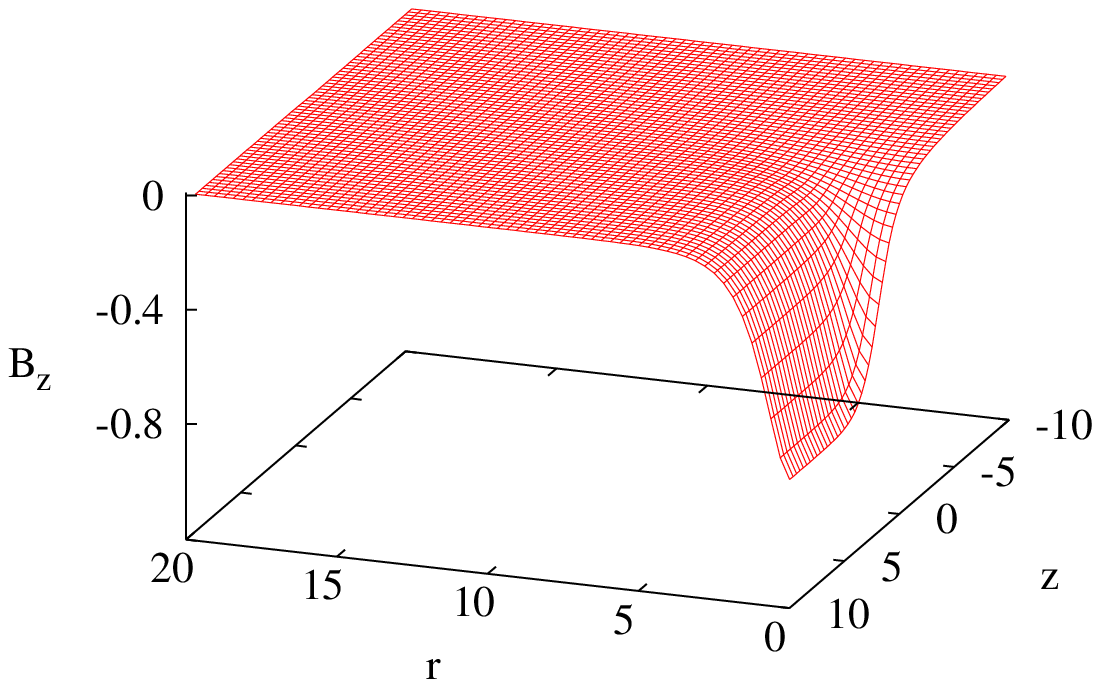}
\caption{Component $B_z$ of the magnetic field.}
\label{Bz}
\end{figure}

\begin{figure}[ht]
\centering
\includegraphics[width=4.5in]{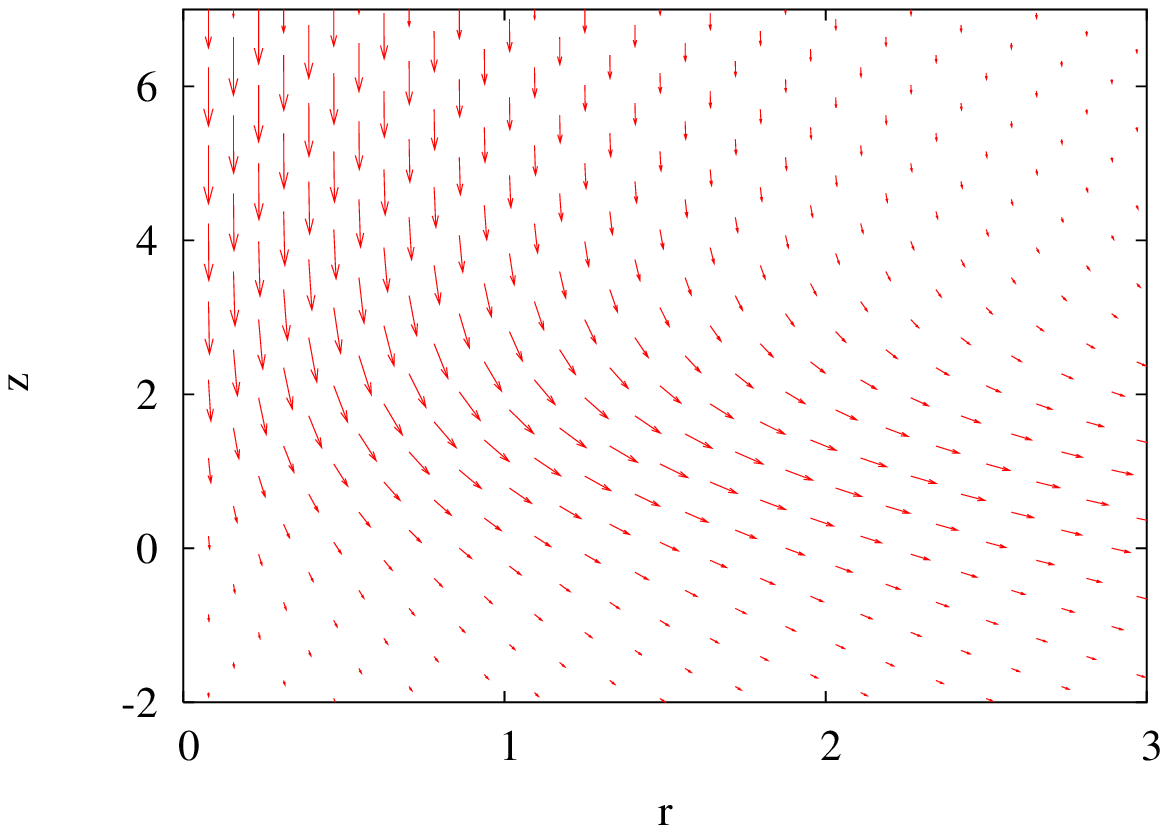}
\caption{2D plot of the magnetic field $\vec B$.}
\label{B}
\end{figure}

\begin{figure}[ht]
\centering
\includegraphics[width=4.5in]{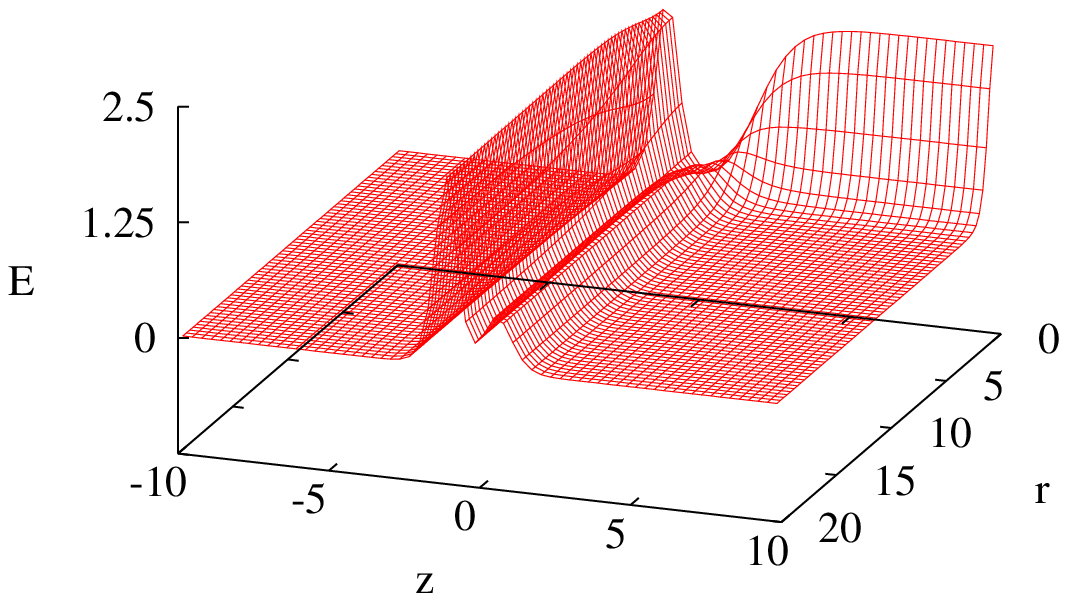}
\caption{Energy density for a local string ending on a wall.}
\label{lsw_en}
\end{figure}

\begin{figure}[ht]
\centering
\includegraphics[width=4.5in]{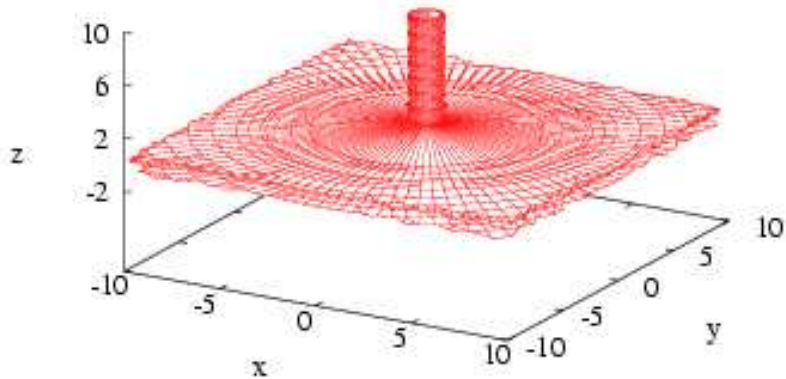}
\caption{Contour plot for the energy density for a local string
ending on a  wall.} \label{le}
\end{figure}

We plot our results as before, with the energy density now given
by
\begin{equation}
\begin{split}
E = T^0{}_0 = -\mathcal{L} &=
\frac12\left(\frac{\partial\phi}{\partial r}\right)^{\!2}+
\frac12\left(\frac{\partial\phi}{\partial z}\right)^{\!2}+
\left(\frac{\partial\psi_2}{\partial r}\right)^{\!2}+
\left(\frac{\partial\psi_2}{\partial z}\right)^{\!2}+\\ &\qquad
\left(\frac{\partial R_1}{\partial r}\right)^{\!2}+
R_1^2\,\left(\frac{1}{r}+eA\right)^{\!2} + \left(\frac{\partial
R_1}{\partial z}\right)^{\!2}+ \frac12\,\vec B^2+V(\phi,R_1,\psi_2) \
.
\end{split}
\end{equation}
We also plot the magnetic field associated with the gauge field of
the string, given by
\begin{equation}
\vec B =
-\frac{\partial A}{\partial z}\,\hat r +
\frac1r\,\frac{\partial(rA)}{\partial r}\,\hat z \ .
\end{equation}
The $\phi$ field domain wall configuration is plotted in
Fig.~\ref{lsw_phi}. The intersection with the $\psi_1$ string at $r=0$
is clearly visible.
The corresponding configuration for $\psi_1$ is shown in
Fig.~\ref{lsw_psi1}. Note the string for $z=+\infty$.
The field $\psi_2$ configuration is plotted in
Fig.~\ref{lsw_psi2}. It has the form of a wall (except for an
interaction with the string for $r=0$ and $z>0$) with half the
height of the Dirichlet wall in the sense that it goes from 0 (at
$z=+\infty$) to $w=1$ (at $z=-\infty$).
Finally the gauge field configuration is shown in Fig.~\ref{A}. We
can verify that $A\to r^{-1}$ as $(r,z)\to+\infty$.
We plotted the $r$ component of the magnetic field, $B_r$, in
Fig.~\ref{Br}.
The $z$ component of the magnetic field, $B_z$, is illustrated in
Fig.~\ref{Bz}
A 2-D plot of the vector field $\vec B$ is shown in Fig.~\ref{B}.
The energy density plotted in Fig.~\ref{lsw_en} illustrates the
string merging at the origin with the Dirichlet wall.
A corresponding contour plot of the energy density in
Fig.~\ref{le} shows the string ending on the wall very clearly.

\null\eject
\null\eject
\null\eject

\acknowledgments We thank Joel Rozowsky for helpful
discussions. MB is supported by the US Department of Energy (DOE)
under contract No.~DE-FG02-85ER40237 and by the National Science
Foundation (NSF) under grant DMR-0219292.  ADF and MT are supported in
part by the NSF under grant PHY-0094122. MT is a Cottrell Scholar of
Research Corporation.

\end{document}